\documentclass[11pt]{book}
\usepackage{graphics}
\usepackage{graphicx}
\usepackage{amssymb,amsmath}
\usepackage{makeidx,landscape}
\usepackage{lscape}
\usepackage[square,sort,comma,numbers]{natbib}
\usepackage[figuresright]{rotating}
\usepackage{supertabular} 
\usepackage{url}
\usepackage{color}
\usepackage{wrapfig}
\DeclareTextCommandDefault{\regcopyright}{\textcircled{\textsc{R}}}

\usepackage{graphicx}

\newcommand{\hs}{\hspace{.2in}}
\newcommand{\hso}{\hspace{0.1in}}

\newcommand{\VT}[1]{\ensuremath{{V_{T#1}}}}

\newbox\sectsavebox
\setbox\sectsavebox=\hbox{\boldmath\VT{xyz}}

\usepackage{natbib}
\usepackage{hyperref,url}

\newcommand\BibTeX{{\rmfamily B\kern-.05em \textsc{i\kern-.025em b}\kern-.08em
T\kern-.1667em\lower.7ex\hbox{E}\kern-.125emX}}

\begin{document}

\raggedbottom

\pagenumbering{arabic}

\setcounter{chapter}{13}
\chapter[Control Challenges]
{Control Challenges}

{\bf \Large Quanyan Zhu} \bigskip \\ \medskip
{Tandon School of Engineering}\\ \medskip
{New York University}\\ \medskip
E-mail: {qz494@nyu.edu}

\bigskip
\bigskip
\bigskip
\bigskip
%
%
%



\addcontentsline{toc}{section}{\hs \hso \, Abstract}
\addcontentsline{toc}{section}{\hs \hso \, Objectives}

\section*{Abstract}

 In this chapter, we introduce methods to address resiliency issues for control systems. The main challenge for control systems is its cyber-physical system nature that strongly couples the cyber systems with physical layer dynamics. Hence, the resiliency issues for control systems need to be addressed by integrating the cyber resiliency with the physical layer resiliency. We introduce frameworks that can provide a holistic view of the control system resiliency and a quantitative design paradigm that can enable an optimal cross-layer and cross-stage design at the planning, operation, and recovery stage of control systems. The control systems are often large-scale systems in industrial application and critical infrastructures. Decentralized control of such systems is indispensable. We extended the resiliency framework to address distributed and collaborative resiliency among decentralized control agents. 

\newpage 

\section*{Objectives}
The objectives of this chapter is to provide an overview of resilient control systems, including understanding:
\begin{enumerate}
\item Cross-layer tradeoffs between security at the cyber layer and resiliency at the physical layer.
\item Cross-stage resiliency design, including ex-ante planning, interim operation, and ex-post recovery. 
\item Games-in-games design paradigm for the multi-stage and multi-layer design of resilient control systems. 
\item Control challenges for distributed control systems.
\end{enumerate}

 \section{Resiliency Challenges in Control Systems}

Modern control systems are equipped with information system technologies (ICTs) that can provide situational awareness of the plant and enable a fast response to emergency and security breaches. As the control systems benefit from the enhanced functionalities and autonomy, the cybersecurity vulnerabilities become prominent issues to resolve. Adversaries can take a sequence of moves and launch multi-phase and multi-stage attacks from the early reconnaissance to the objective of data exfiltration. This structure of attacks is known as the cyber kill chain. The defense against such attacks includes detection of an adversary, disruption of the network system, cyber deception to create uncertainties and costs for the attacker, and many other techniques (see \cite{jajodia2011moving,zhu2013game,pawlick2019game,jajodia2016cyber}). Despite the effort in developing cyber defense for control systems, the perfect security is not always achievable. Achieving perfect security would either require a cost-prohibitive amount of resources to maintain the security or lead to a degradation of system operability and usability. Hence, it is important to shift the focus from security-centered design to the paradigm of secure and resilience design of control systems. Adding resiliency as an additional dimension to the new paradigm complements the sole reliance on ex-ante perfect security technologies as a solution with interim and ex-post resiliency mechanisms as solutions when the ex-ante security mechanism fails to protect the control system. 

Resiliency is a key system concept that focuses on the post-event behaviors of a system. For example, when a cyber attack has successfully reached its target, any security mechanism becomes futile at this point.  Reliance should be on resiliency mechanisms that can reduce the impact of this successful attack and enable a fast recovery to restore the operations to their normal state. On the contrary, security mechanisms are often used as tools to prevent successful attacks or attackers from achieving their objectives. Hence security and resilience are dual concepts and there are fundamental relationships between them. 

First, there exists a tradeoff between security and resiliency. When there is perfect security, resiliency becomes unnecessary. The need for resiliency is high when the security is poor.  Second, security and resiliency resolutions have to be implemented jointly. One cannot count on either security or resiliency solutions to safeguard the system from adversarial behaviors.  The cost for security and resiliency solutions are often different. The level of security and resiliency implemented in the system should be  optimized together.

Understanding these relationships provides a fundamental understanding toward designing secure and resilient systems. There are two key challenges to design secure and resilient systems. One is the cross-layer challenge and the other is the cross-stage challenge. Cross-layer challenges refer to the fact that the security issues often reside at the cyber layer of the system while the resiliency often deals with the last-mile issue and in the context of control systems, the resiliency issues sit at the physical layer of the systems. In other words, the failure in the protection at the cyber layer can lead to malfunctioning of the physical system performance. The joint design of secure and resilient solutions is naturally cross-layered. It would require  understanding the dependency and interdependency among human, cyber, and physical layers of the control system. For example, the failure of cyber defense against an advanced persistent threat would lead to data injection on sensors and the manipulation of the controller of the physical plant. Human negligence in system configurations would lead to cyber vulnerabilities that can be exploited by the attackers to reach targeted physical assets. 

The second challenge of resilient control systems arises from the fact that resiliency is a dynamical system concept. Improving resiliency involves multi-stage planning and design, including ex-ante planning, interim execution, and ex-post recovery. The ex-ante stage refers to the planning stage before the control system starts to run. At this stage, one needs to invest resources and plan contingencies to provide information and physical resources to enable fast recovery at later stages. For example, one can add redundancies such as sensors and power generators to prepare for an attack on sensors and attack-induced loss of power. One can also design secure and switching controllers in advance to prepare for the worst-case operational environment and provide a contingency controller when the control system encounters failures. The offline design of such controller at the ex-ante stage prepares for a set of anticipated attacks in later stages.  However, unanticipated attacks can still occur. In addition, the preparation for a large set of events can be expensive. There is a tradeoff between what events should be prepared for at the ex-ante stage and what events should be dealt with in later stages. Hence, from the set of anticipated events,  the high-impact and high-frequency events should be selected first.

\begin{figure}[ht]
\centering
\includegraphics[width=0.9\textwidth]{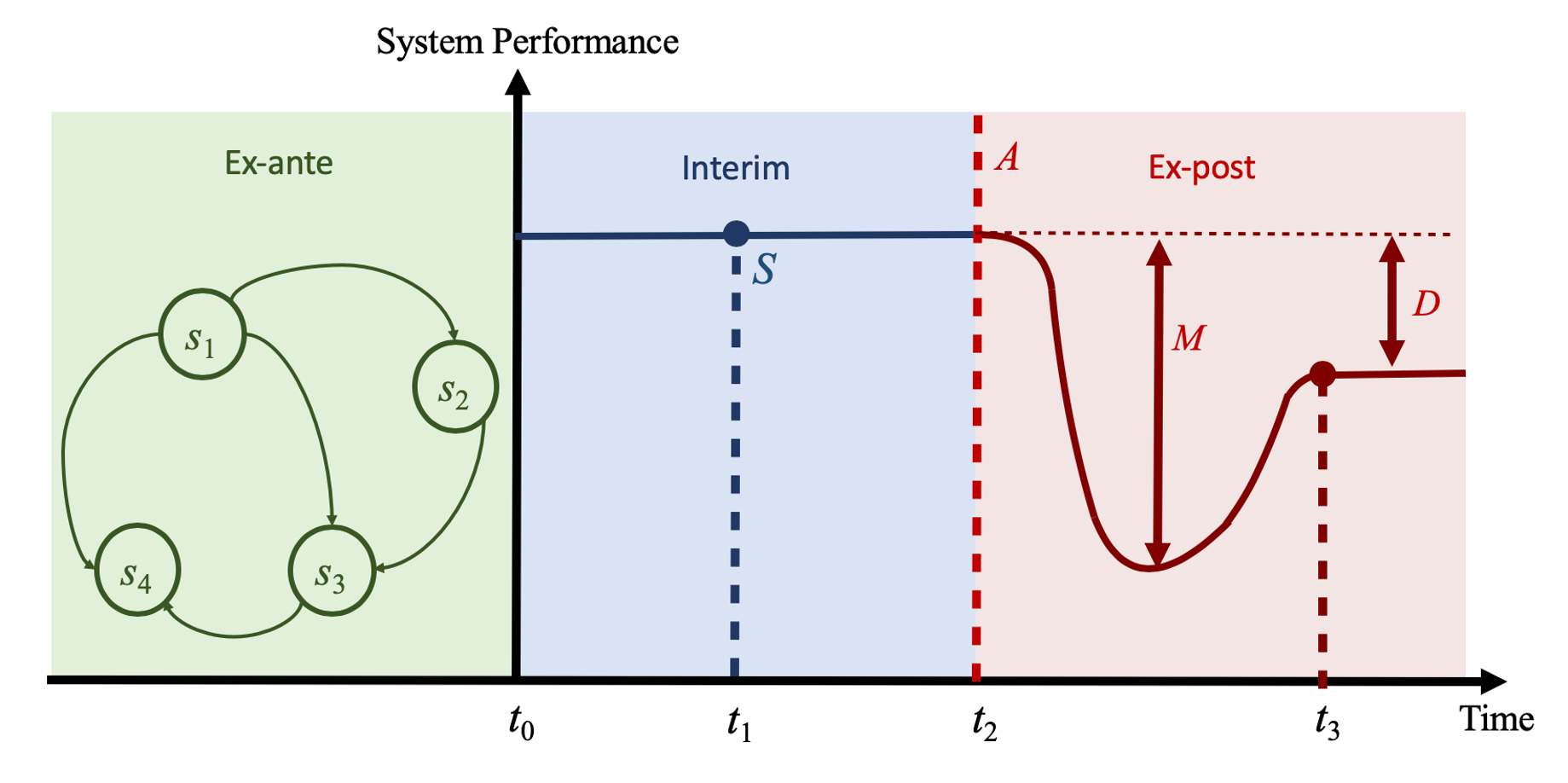}
\caption{Multi-stage planning and design for resilient control systems: ex-ante stage plans resiliency. Interim stage executes resilient control. Ex-post stage restores system performance after unanticipated attacks. }\label{cross-stage} 
\end{figure}

The interim stage refers to the operation stage of the control systems. By running the controller that prepares for a selected set of anticipated events, the control system can run smoothly and intelligently when it encounters these prepared events.  The control system can switch to a different control logic or leverage built-in resources to handle events that occurred. In Figure \ref{cross-stage}, at time $t_1$, an anticipated event has occurred; for example, there is a loss of a sensor. By switching to a redundant sensor, the interim stage operation does not suffer any loss of system performance. 

The ex-post stage refers to the stage where unanticipated event occurs and the recovery process kicks off. The probability of running into an unanticipated event depends on the set of events prepared for by the ex-ante controller. In Figure \ref{cross-stage}, the unanticipated event $a$ occurs at $t_2$. The expected value of $t_2$ would depend on the frequency of Event $A$ and whether the ex-ante stage has been taken into account. Since no contingency plan has been made for Event $A$, the system will suffer a performance degradation. The ex-post resiliency design aims to detect the event quickly and find a self-healing mechanism that can restore the system to its normal operation or an acceptable performance level. Depicted in Figure \ref{cross-stage}, the system performance degrades to its maximum difference $M$ and then gradually recovers to the level $D$ at time $t_3$. The ex-post resiliency is measured by the total performance loss after the event. The goal of ex-post resiliency is to minimize this loss by responding fast to the event and developing an effective restoration plan. The restoration plan would depend on the configurable resources available to the system and the ex-ante resource investment made for the control system. For example, when an attack has successfully compromised a centrifuge control system, the system can detect and reboot. The cost for reboot depends on whether the control system is well equipped with computational resources and human resources to enable the fast response.

It is clear that the concept of resiliency spans three stages of the control systems. The planning of ex-ante resiliency affects the resource availability of and the need for ex-post resiliency. Hence, the ex-ante resiliency planning and the ex post resiliency mechanism have to be jointly designed. 
This chapter will introduce a games-in-games framework to develop a cross-layer cross-stage design framework. We will use examples from unmanned vehicles to illustrate the games-in-games framework.

\section{Resiliency Design Framework}

\begin{figure}[ht]
\centering
\includegraphics[width=0.7\textwidth]{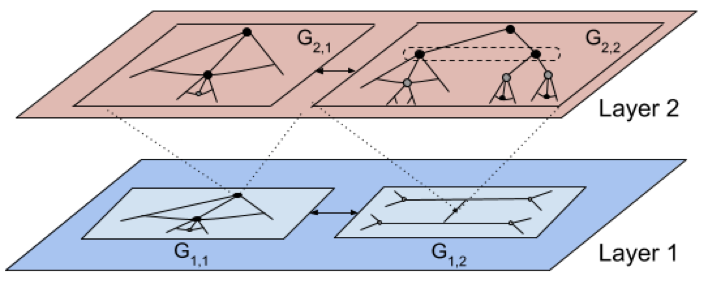}
\caption{Composition of cyber games and physical layer games together to form a cross-layer adversarial model. }\label{cross-layer} 
\end{figure}

A natural framework to enable the cross-layer design is to use the games-in-games framework introduced in \cite{zhu2015game}. Game-theoretic methods have been used to capture different cyber-attack scenarios and models including jamming (see \cite{zhu2011eavesdropping,zhu2011eavesdropping,song2019performance,zhu2010stochastic}), spoofing (see \cite{zhang2017strategic,xu2015cyber}), and network configurations (see \cite{zhu2011indices,zhu2010network,huang2020dynamic,huang2019adaptive,rass_gadapt:_2016}). Interested readers can refer to \cite{manshaei2013game} and \cite{pawlick2019game} for recent surveys. Each game can be used to model a cyber-attack scenario. Composing these games together forms a set of anticipated adversarial behaviors at the cyber layer to be considered. Similarly, the attack behaviors at the physical layer can also be modeled by a set of physical layer games. For example, the attacker can choose to inject bad data into sensors while the defender can choose which sensors to use. The defender can determine the switching policy of the controllers while the attacker can determine how to compromise the logic of one of the control laws. Composing the physical layer games together provides a framework to design ex-ante and ex-post resiliency mechanisms. The cyber games and the physical layer games are interdependent. The outcome of one game would lead to a new game. The games-in-games framework is illustrated in Figure \ref{cross-layer}. Cyber games $G_{2,1}$ and $G_{2,2}$ are composed together to form $G_2$. The physical layer games $G_{1,1}$ and $G_{1,2}$ are composed to form $G_1$. $G_1$ and $G_2$ are interconnected to form a larger game $G$.

A natural control framework to capture the multi-stage features of resilient control systems is model-predictive control or moving-horizon techniques. The moving-horizon control looks $N$ steps into the future, prepares for possible events, and finds optimal control strategies to be implemented at the current stage. The ex-ante control design takes into account a credible set of anticipated adversarial models. This design process can be formulated as a game between the control designer and adversaries. The game model can be chosen to capture the cross-layer nature of the control system as illustrated in \\ 
 Figure \ref{cross-layer}. The ex-ante control is implemented immediately after the design. If no unanticipated behaviors occur, this game-based moving-horizon control continues. When unanticipated adversarial behavior happens in the next moving horizon, the control design not only prepares for the anticipated adversarial behaviors but also determines ways to recover from the unanticipated attack. The moving-horizon technique directly incorporates the ex-ante planning by looking into possible future events and the ex-post recovery by immediately reconfiguring the system in the next moving horizon.

\subsection{Control of Autonomous Systems in Adversarial Environment}\label{uavcontrol}

To illustrate the moving-horizon resilient control design paradigm, we present a case study of control for mobile autonomous systems in adversarial environment. The objective of the control is to maintain the connectivity of the autonomous system in an environment where an attacker can jam the communications among the autonomous systems. The operator does not know the capability of the attacker (i.e., how many links the attacker can jam, and where and when the attacker will jam the communications). The operator prepares for the anticipated level of attacks and plans the control in a moving-horizon way. At every time $k$, the operator solves the following problem:
\begin{equation}\label{QK}
Q^k: \ \ \ \ \ \ \max_{x(k+c)} \min_{e \in E} \lambda_2(e, x(k+c)).
\end{equation} 

Here $x(k)$ is the configuration of the mobile network at time $k$ (i.e., the position of the mobile agents). Two mobile agents can form a link when they are sufficiently close within a desirable range of communications.  Hence, the configuration $x(k)$ includes a network whose connectivity is described by the algebraic connectivity of the network, denoted by $\lambda_2$ (i.e., the second-smallest eigenvalue of the associated Laplacian matrix). At each time step $k$, the operator determines where the agents should move to in the next time step $x(k+c)$, where $c$ is a time interval. The control is constrained by the physical dynamics of autonomous systems. To maintain connectivity, the operator aims to maximize the level of connectivity as much as possible at time $k+c$ by anticipating the worst-case adversarial behaviors given a certain level of attacks, described by the set $E$. For example, the operator can anticipate one-link removal and determine how agents can move and maintain connected secure to such one-link removal. On one hand, when the capability of the attacker is higher, it becomes more difficult for the operator to find a controller to be secure to the attack level. On the other hand, if the operator underestimates the attacker, his control strategies will not achieve a desirable connectivity. Hence, the operator should anticipate a reasonable level of attacks and design an ex-ante controller that will be implemented at the interim stage. When the attacker's capability does not exceed the attack level anticipated by the ex-ante controller, the network connectivity is maintained as expected. When the attacker's capability exceeds the anticipated level, the network connectivity may not be achieved and the ex-post resiliency mechanism will be designed at time $k+c$. In other words, the problem for the operator at $k+c$, $Q^{k+c}$ includes how to react to the failure of network connectivity and how to design new control laws to heal the broken links while anticipating new attacks at time $k+c$. The anticipation of attack levels can be adjusted from time $k$ to $k+c$. 

\begin{figure}[ht]
\centering
\includegraphics[width=0.7\textwidth]{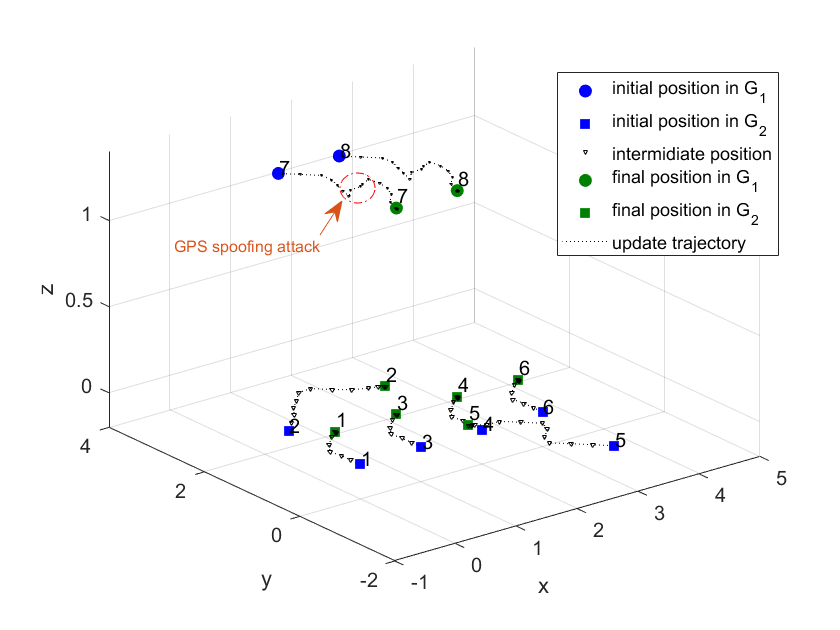}
\caption{Network connectivity: dynamic configuration of secure robotic network. The GPS spoofing attack is introduced at time Step 9, and it lasts for 5 steps. }\label{trajectory} 
\end{figure}

\begin{figure}[ht]
\centering
\includegraphics[width=0.6\textwidth]{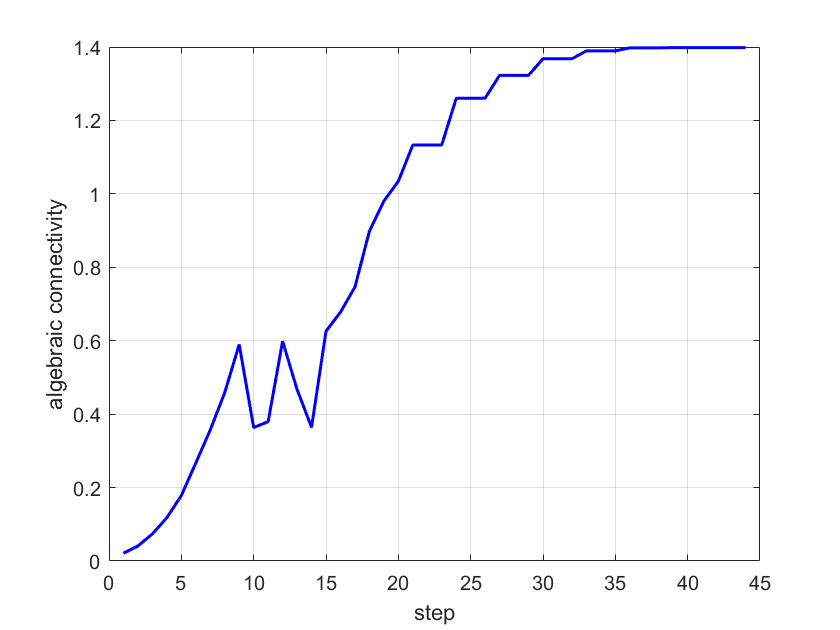}
\caption{The network connectivity over time.}\label{connectivity} 
\end{figure}

This moving-horizon framework has been shown to be effective in obtaining the self-adaptability, self-healing, and resilience of the Internet of Battlefield Things (IoBTs) (see \cite{Chen2019optimal,farooq2018secure,chen2019control}). Specifically, the unmanned ground vehicle (UGV) network should coordinate its actions with the unmanned aerial vehicle (UAV) network and the soldier network to achieve a highly connected global network. The designed decentralized algorithm yields an intelligent control of each agent to respond to others to optimize real-time connectivity under adversaries. Figure \ref{trajectory} shows an example of a two-layer robotic network that is robust to jamming attack at every step. Figure \ref{connectivity} shows the algebraic connectivity over time associated with the two-layer network. Furthermore, the agents can respond to the spoofing attack quickly, which shows the resilience of the control strategy. The developed moving-horizon framework can be further adopted to address the mosaic control design as the framework provides built-in security and resilience for each component in the system, which guarantee the performance of the integrated system.

\subsection{Cross-Layer Defense for Cloud-Enabled Internet of Controlled Things}\label{ioctgames}

To illustrate games-in-games framework, we consider the Internet of Controlled Things (IoCT) that integrates computing, control, sensing, and networking. The IoCT relies on local clouds to interface between heterogeneous components. The cloud-enabled IoCT is composed of three interacting layers: a cloud layer, a communication layer, and a physical layer. In the first layer, the cloud-services are threatened by attackers capable of APTs and defended by network administrators (or ``defenders").
The interaction at each cloud-service is modeled using the {FlipIt} game recently proposed by \cite{bowers2012defending} and  \cite{van2013flipit}.  We use one FlipIt game per cloud-service. In the communication layer, the cloud services, which may be controlled by the attacker or defender according to the outcome of the FlipIt game--transmit information to a device that decides whether to trust the cloud-services. This interaction is captured using a signaling game. At the physical layer, the utility parameters for the signaling game are determined using optimal control. The cloud, communication, and physical layers are interdependent. This motivates an overall equilibrium concept called Gestalt Nash equilibrium (GNE). GNE requires each game to be solved optimally given the results of the other games. Because
this is a similar idea to the concept of best response in Nash equilibrium, we call the multi-game framework a game-of-games.

\begin{figure}[ht]
\centering
\includegraphics[width=0.55\textwidth]{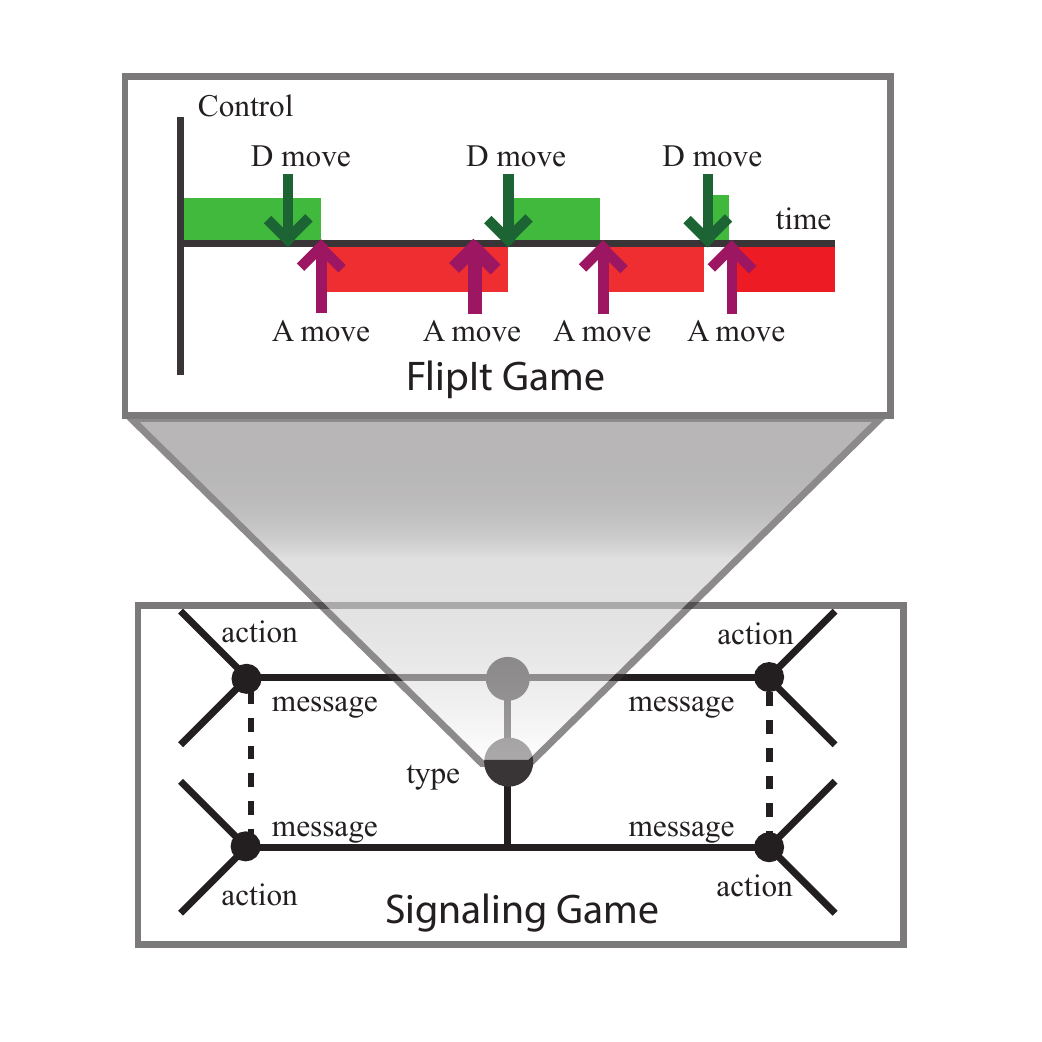}
\caption{Conceptual model of the composed $G_1$ and $G_2$. }\label{twoGameDia} 
\end{figure}

\begin{figure}[h]
\centering
\includegraphics[width=0.6\textwidth]{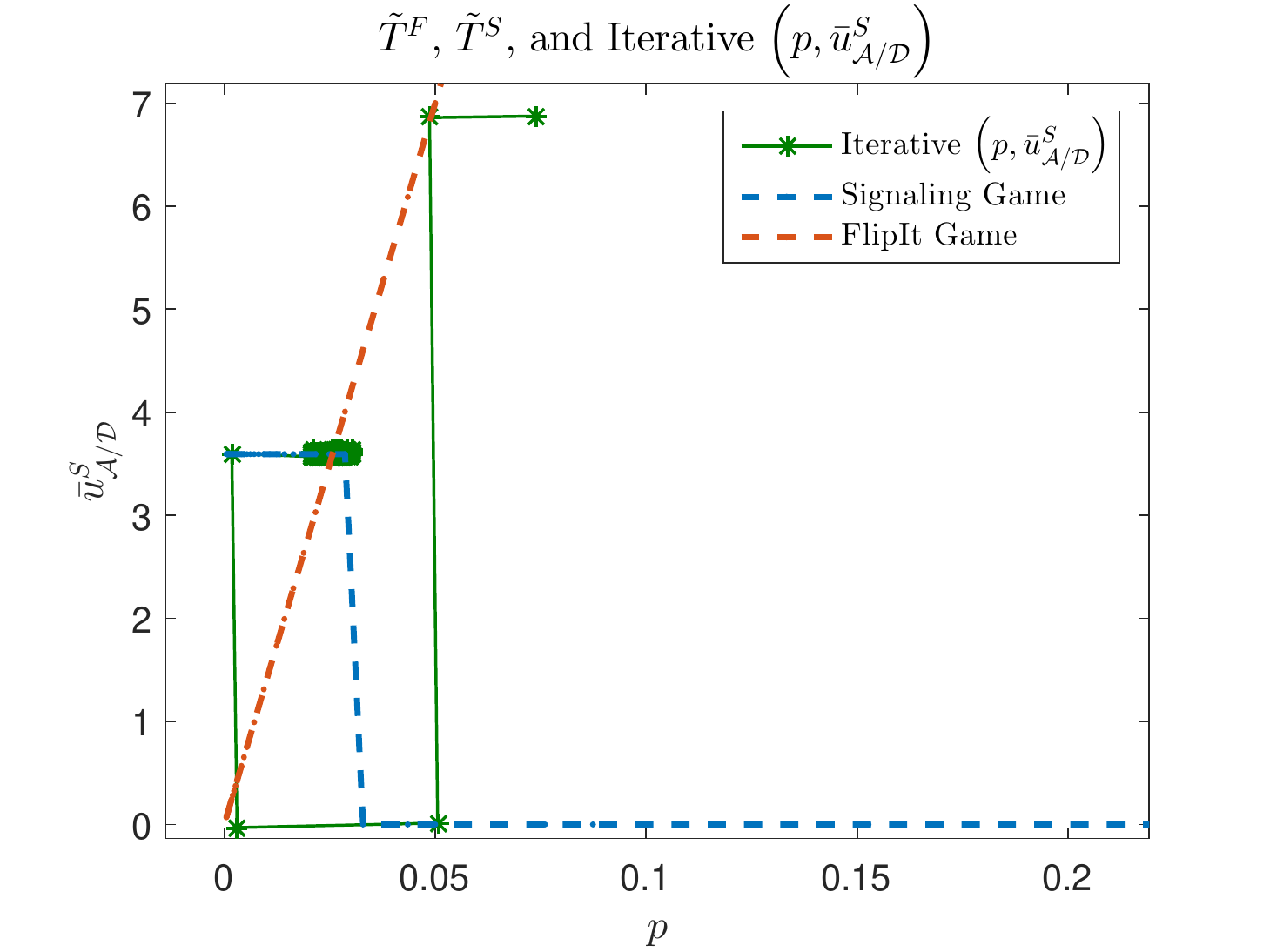}
\caption{Iterations to compute a Gestalt Nash equilibrium.}\label{severalIterStable}
\end{figure}

A composition of a {FlipIt} game $G_1$ (e.g., \cite{van2013flipit,bowers2012defending}) and a signaling game $G_2$ (e.g., \cite{spence1973job,banks1987equilibrium}), depicted in Figure \ref{twoGameDia}, has been used to provide a strategic trust management in Internet of things (IoT) networks vulnerable to advanced persistent threats.  The game $G_1$ describes the strategic interactions between an attacker and a cloud service provider where the attacker aims to gain control of the computing resources and the cloud service provider protects and audits the system. The game $G_2$ describes the information asymmetry between the sender of the message (i.e., the computational results) and the IoT as the receiver of the message. $G_1$ and $G_2$ are composed sequentially as in Figure \ref{twoGameDia}.
An attacker and a defender play the {FlipIt} game for control of the cloud. Then, the winner sends a command to the device in the signaling game.  The Gestalt Nash equilibrium of the meta-game predicts the risk of sequential adversarial interactions. As shown in Figure \ref{severalIterStable}, the equilibrium can be computed as the intersection of blue and red curves in an iterative manner within a finite number of steps.  

\section{Resiliency for Decentralized Control Systems}

Resilient design for distributed systems requires engineering agents with flexible interoperability and the capability of self-adaptability, self-healing, and resiliency. It is important that systems can achieve its objective when one node goes away or fails. In addition, a subsystem can respond to other subsystems in a non-deterministic/stochastic way.  Such a design increases the composability and modularity of the system design. For example, agents can randomly arrive and respond in a stochastic but structured way to other agents in an uncertain environment. However, the structured randomness leads to emerging system behaviors that manifest desirable properties for the objective of the entire mission.

Systems that have such properties are easily composable and resilient-by-design. Without a pre-planned integration among agents, the agents can adapt their responses and reconfigure their own systems based on the type of agents with whom they interact. Agents can be easily composed to achieve a prescribed objective through an unprescribed path. In adversarial environments, the agents can reconfigure their responses and roles to achieve the global mission in spite of the failures of nodes and links. The system can still operate when one piece is missing. It is the epicenter of the mosaic designs.

We can leverage the games-in-games principle as discussed in \cite{zhu2015game} and \cite{huang2019dynamic}  to provide a theoretical underpinning and a guideline for distributed resilient control designs. The games-in-games approach integrates three layers of design for each agent: strategic layer, tactical layer, and mission layer. At the strategic layer, the agents learn and respond to their environment quickly to unanticipated events such as attacks, disruptions, and changes of other agents. At the tactical layer, the agents plan for a more extended period of time by taking into account the long-term interactions with the environment and other agents. The agents can make goal-oriented planning at each stage. At the mission layer, the agents develop stage-by-stage planning of multi-stage objectives to achieve the mission despite the uncertainties and online changes.

Each layer corresponds to a game of a different scale. For example, at the tactical layer, a game associated with an agent describes its interaction with an adversary (e.g., a jammer, a spoofer, or a sudden loss of a neighboring node). Solutions to this game can prepare nodes for unanticipated attacks and secure the agents. At the strategic layer, an $N$-person dynamic game describes the longer-term interactions among cooperative agents, each seeking control policies to achieve individual stage objectives. The individual control would lead to achieving global objectives (such as connectivity and network formation). At the mission layer, each agent plans their stage objectives for each stage at the tactical layer. This planning is obviously under a lot of uncertainties and needs to be achieved in a moving-horizon way.
 
The games-in-games framework enables security and resilience by design. From the perspective of security, the framework anticipates the attack behavior and designs a control policy that would prepare to defend against the anticipated attacks. The framework provides a clean-slate design and built-in security for each system component that would lead to the security of the integrated system. 
 From the perspective of resiliency, the framework enables each system to respond to the unanticipated events at each time instant. Each agent can respond to events that inflict damages on the agent and go through a self-healing process that can recover itself from the attacks and failures if possible. If the full recovery is not achievable, the agents will develop control strategies that will allow a graceful performance degradation.

\section{Conclusions and New Challenges}

This chapter has introduced two key challenges for resilient control systems. One is the cross-layer challenge requiring an integrated cyber-physical perspective.  The optimal design pivots on the understanding of the tradeoffs between the security at the cyber layer and the resiliency at the physical layer. The other is the cross-stage resiliency mechanism that requires three stage resiliency design including ex-ante planning, interim operation, and ex-post recovery. The games-in-games framework provides a design paradigm for the multi-stage and multi-layer design of resilient control systems. The framework is particularly useful for distributed control systems by designing modular agents that can work together. 

The chapter has presented two applications. One is the control of autonomous vehicles in adversarial environment and the other one is the cross-layer defense for cloud-enabled IoCTs. The methodology presented in these two case studies can be further generalized and made applicable to other systems, including smart grids, transportation systems, and  manufacturing systems. The new challenge with these large-scale systems is the scalability of the solutions and the incompleteness of the situational awareness. The resiliency design would need mechanisms that can achieve scalable yet suboptimal resiliency and deal with time delay and non-global system state information.

\section{Further Reading}

Game-theoretic methods have been widely used to model adversarial behaviors in wireless communications (See \cite{zhang2017strategic,zhu2012interference,zhu2011dynamic,zhu2010stochastic,farhang2015phy,zhu2011eavesdropping,zhu2010dynamic}), network configuration (See \cite{fung_facid:_2016,manshaei2013game,zhu2012guidex,Quanyan2010ACC}), and control systems (See \cite{zhu2015game,rieger2019,Quanyan2011ECDC,Quanyan2013CCPS,miao2018hybrid,xu2015cyber}). The applications of security games have addressed the critical infrastructure protection (See \cite{HuangAPT,zhu2018multi,chen2017dynamic,chen2017interdependent,RassZhuBook2020}), cyber insurance (See \cite{zhang2019flipin,zhang2017bi,zhang2016attack,chen-allerton}), cyber deception (See \cite{pawlick2018modeling,zheng2012dynamic,zhu2012deceptive,pawlick2019game,huang2020dynamic,horak2017manipulating}), adversarial machine learning (See \cite{huang2019deceptive,pawlick_mean-field_2017,zhang2015secure,zhang2018game}), and network systems (See \cite{chen_2019_TIFS,chen-TCNS-19-optimal,chen-tifs-17,8779673,pawlick2017strategic}).

Game-theoretic approaches have also addressed the resilient design of control systems. Games-in-games approach has been proposed  
 Section \ref{uavcontrol} is based on the work of \cite{chen2019control,chen-CDC-16} in which two layers of mobile agents are controlled to maintain network connectivity. Section \ref{ioctgames} is based on the work of \cite{pawlick2015flip,pawlick_TIFS_18}. The Internet of Controlled Things (IoCTs) extends the concept of Internet of Things (IoTs) and studies a cloud-enabled sensing and actuation architecture with three layers of interacting systems including the cloud, communication networks, and sensor-actuator networks. The games-in-games principle was first discussed in \cite{zhu2015game} and has been used for addressing applications of multi-layer networks (See \cite{Chen2019optimal,nugraha2019subgame,xu2018cross,xu2016cross,xu2017game,xu2015cyber}) and trust management (See \cite{pawlick_TIFS_18,pawlick2015flip,pawlick2017strategic}). The recent book \cite{zhucross} has provided a comprehensive introduction to game- and control-theoretic techniques for cross-layer cyber-physical systems.

One important application of resilient control systems is critical infrastructures, including power grids, transportation, and manufacturing systems. They are often legacy systems that modern ICT technologies build on to improve their efficiency and expand their functionalities. Due to the large-scale nature of  critical infrastructures, their resilience has to be designed in a decentralized manner. It is critical to understand and capture the interconnections between components within the infrastructure as well as the interdependencies among different infrastructures. Network modeling and design approaches (e.g., \cite{Zimmerman-17,zimmerman2018network,zimmerman_promoting_2016,rinaldi2001identifying}) have been used to provide a holistic framework for understanding cyber-physical interdependencies in infrastructure systems.
Readers can refer to \cite{chen2019game,peng2020distributed,huang2018factored,huang2017large,huang2018distributed} for further materials on the topic of resilient interdependent infrastructures.

Resiliency for decentralized systems is a fundamental challenge for increasingly complex and connected systems. Recent works \cite{Chen2019optimal,chen-TCNS-19-optimal,chen2017dynamic,chen2019dynamic,chen2017heterogeneous} have addressed this issue from the perspective of network science. They have provided a design methodology to create networks that are resistant to link removals. In particular, \cite{chen2019dynamic} has provided a trade-off analysis between the ex-ante robustness  and the ex-post recovery of the network. This framework has also been recently extended to the context of multi-agent systems, where multiple mobile agents communicate locally to accomplish a mission (e.g., rendezvous and formation). Interested readers can refer to \cite{nugraha2019subgame,nugraha2020dynamic,nugraha2020dynamic,juntaosmartgrid,chen2017stackelberg,chen2015resilient} for recent development on this topic.

Chapter 12 of this book introduces Grid Game, a simulation and learning platform for the electric grid and resilient controls. It is a useful tool to simulate and understand the multi-agent behaviors described in this chapter in the context of electric power systems. For example, 
a multiplayer microgrid game has been formulated in \cite{maharjan2013dependable} and its related extensions to multi-layer games have been introduced in \cite{chen2017stackelberg,maharjan2015demand,juntaosmartgrid}. The decentralized architecture of the multiplayer microgrids improves the resiliency of local microgrids when they are subject to attacks or natural failures. Readers can refer to Problem 14.3 for an exercise problem that builds on the Grid Game platform.

\section{Thoughtful Questions to Ensure Comprehension}

\noindent {\bf Problem 14.1} Consider a smart electric power system, discuss the cross-layer control challenges of the cyber-enabled power systems. Present methods to improve the resiliency of the power systems. Discuss how to scale the resiliency solutions when the size distribution and transmission network increases. \textit{Hint: The readers can refer to the case studies in \cite{zhu2015game}.}

  \bigskip
\noindent {\bf Problem 14.2} 
Write down the dynamics of the unmanned  aerial vehicles (UAVs) and create a case study of a network of five UAVs. Simulate three steps by solving  a sequence of three optimization problems (\ref{QK}). Use MATLAB$^{\tiny{\regcopyright}}$ to present the simulation results. {\it Hint: The complete description of the algorithm can be found in \cite{chen2019control}.}

  \bigskip
\noindent {\bf Problem 14.3} 
In this problem, we will investigate the game-theoretic algorithms for distributed energy systems. We consider an energy system that consists of multiple microgrids that are modeled as self-interested players that can operate, communicate, and interact autonomously
to efficiently deliver power and electricity to their consumers. Consider a two-microgrid case where each of them aims to minimize the cost of production and maximize the quality of the power quality. The utility function of this game has been described in \cite{juntaosmartgrid}. Use GridGame introduced in Chapter 12 to create a simulation platform of two players. Implement the distributed algorithm in (11) in \cite{juntaosmartgrid} for each microgrid and observe the behavior of the grid. Discuss the resiliency of the grid under the following two scenarios: (a) there is a false data injection attack on the  PMU angle measurement; (b) One microgrid suddenly fails and cannot generate power to meet the demand.  \textit{Hint: The readers can refer to the case studies in Section IV-B of \cite{juntaosmartgrid}.}


 \bibliographystyle{abbrv}
\bibliography{Wiley}%

\begin{thebibliography}{10}

\bibitem{banks1987equilibrium}
J.~S. Banks and J.~Sobel.
\newblock Equilibrium selection in signaling games.
\newblock {\em Econometrica: Journal of the Econometric Society}, pages
  647--661, 1987.

\bibitem{bowers2012defending}
K.~D. Bowers, M.~van Dijk, R.~Griffin, A.~Juels, A.~Oprea, R.~L. Rivest, and
  N.~Triandopoulos.
\newblock Defending against the unknown enemy: Applying flipit to system
  security.
\newblock In {\em Decision and Game Theory for Security}, pages 248--263.
  Springer, 2012.

\bibitem{chen2017dynamic}
J.~Chen, C.~Touati, and Q.~Zhu.
\newblock A dynamic game analysis and design of infrastructure network
  protection and recovery: 125.
\newblock {\em ACM SIGMETRICS Performance Evaluation Review}, 45(2):128, 2017.

\bibitem{chen2017heterogeneous}
J.~Chen, C.~Touati, and Q.~Zhu.
\newblock Heterogeneous multi-layer adversarial network design for the
  iot-enabled infrastructures.
\newblock In {\em GLOBECOM 2017-2017 IEEE Global Communications Conference},
  pages 1--6. IEEE, 2017.

\bibitem{chen2019dynamic}
J.~Chen, C.~Touati, and Q.~Zhu.
\newblock A dynamic game approach to strategic design of secure and resilient
  infrastructure network.
\newblock {\em IEEE Transactions on Information Forensics and Security},
  15:462--474, 2019.

\bibitem{Chen2019optimal}
J.~Chen, C.~Touati, and Q.~Zhu.
\newblock Optimal secure two-layer iot network design.
\newblock {\em IEEE Transactions on Control of Network Systems}, 7(1):398--409,
  2019.

\bibitem{chen-TCNS-19-optimal}
J.~Chen, C.~Touati, and Q.~Zhu.
\newblock Optimal secure two-layer {IoT} network design.
\newblock {\em IEEE Transactions on Control of Network Systems}, 2019.

\bibitem{chen2015resilient}
J.~Chen, L.~Zhou, and Q.~Zhu.
\newblock Resilient control design for wind turbines using markov jump linear
  system model with l{\'e}vy noise.
\newblock In {\em 2015 IEEE International Conference on Smart Grid
  Communications (SmartGridComm)}, pages 828--833. IEEE, 2015.

\bibitem{chen-CDC-16}
J.~Chen and Q.~Zhu.
\newblock Resilient and decentralized control of multi-level cooperative mobile
  networks to maintain connectivity under adversarial environment.
\newblock In {\em IEEE Conference on Decision and Control (CDC)}, pages
  5183--5188, 2016.

\bibitem{juntaosmartgrid}
J.~Chen and Q.~Zhu.
\newblock A game-theoretic framework for resilient and distributed generation
  control of renewable energies in microgrids.
\newblock {\em IEEE Transactions on Smart Grid}, 8(1):285--295, 2017.

\bibitem{chen2017interdependent}
J.~Chen and Q.~Zhu.
\newblock Interdependent strategic cyber defense and robust switching control
  design for wind energy systems.
\newblock In {\em Power \& Energy Society General Meeting, 2017 IEEE}, pages
  1--5. IEEE, 2017.

\bibitem{chen-tifs-17}
J.~Chen and Q.~Zhu.
\newblock Security as a service for cloud-enabled internet of controlled things
  under advanced persistent threats: a contract design approach.
\newblock {\em IEEE Transactions on Information Forensics and Security},
  12(11):2736--2750, 2017.

\bibitem{chen2017stackelberg}
J.~Chen and Q.~Zhu.
\newblock A stackelberg game approach for two-level distributed energy
  management in smart grids.
\newblock {\em IEEE Transactions on Smart Grid}, 9(6):6554--6565, 2017.

\bibitem{chen-allerton}
J.~Chen and Q.~Zhu.
\newblock A linear quadratic differential game approach to dynamic contract
  design for systemic cyber risk management under asymmetric information.
\newblock In {\em 2018 56th Annual Allerton Conference on Communication,
  Control, and Computing (Allerton)}, pages 575--582. IEEE, 2018.

\bibitem{chen2019control}
J.~{Chen} and Q.~{Zhu}.
\newblock Control of multi-layer mobile autonomous systems in adversarial
  environments: A games-in-games approach.
\newblock {\em IEEE Transactions on Control of Network Systems}, pages 1--1,
  2019.

\bibitem{chen2019game}
J.~Chen and Q.~Zhu.
\newblock {\em A Game-and Decision-Theoretic Approach to Resilient
  Interdependent Network Analysis and Design}.
\newblock Springer, 2019.

\bibitem{chen_2019_TIFS}
J.~Chen and Q.~Zhu.
\newblock Interdependent strategic security risk management with bounded
  rationality in the {Internet} of things.
\newblock {\em IEEE Transactions on Information Forensics and Security},
  14(11):2958 -- 2971, 2019.

\bibitem{farhang2015phy}
S.~Farhang, Y.~Hayel, and Q.~Zhu.
\newblock Phy-layer location privacy-preserving access point selection
  mechanism in next-generation wireless networks.
\newblock In {\em Communications and Network Security (CNS), 2015 IEEE
  Conference on}, pages 263--271. IEEE, 2015.

\bibitem{farooq2018secure}
M.~J. Farooq and Q.~Zhu.
\newblock On the secure and reconfigurable multi-layer network design for
  critical information dissemination in the internet of battlefield things
  (iobt).
\newblock {\em IEEE Transactions on Wireless Communications}, 17(4):2618--2632,
  2018.

\bibitem{fung_facid:_2016}
C.~J. Fung and Q.~Zhu.
\newblock {FACID}: {A} trust-based collaborative decision framework for
  intrusion detection networks.
\newblock {\em Ad Hoc Networks}, 53:17--31, 2016.

\bibitem{horak2017manipulating}
K.~Hor{\'a}k, Q.~Zhu, and B.~Bo{\v{s}}ansk{\`y}.
\newblock Manipulating adversary's belief: A dynamic game approach to deception
  by design for proactive network security.
\newblock In {\em International Conference on Decision and Game Theory for
  Security}, pages 273--294. Springer, 2017.

\bibitem{huang2017large}
L.~Huang, J.~Chen, and Q.~Zhu.
\newblock A large-scale markov game approach to dynamic protection of
  interdependent infrastructure networks.
\newblock In {\em International Conference on Decision and Game Theory for
  Security}, pages 357--376. Springer, 2017.

\bibitem{huang2018distributed}
L.~Huang, J.~Chen, and Q.~Zhu.
\newblock Distributed and optimal resilient planning of large-scale
  interdependent critical infrastructures.
\newblock In {\em 2018 Winter Simulation Conference (WSC)}, pages 1096--1107.
  IEEE, 2018.

\bibitem{huang2018factored}
L.~Huang, J.~Chen, and Q.~Zhu.
\newblock Factored markov game theory for secure interdependent infrastructure
  networks.
\newblock In {\em Game Theory for Security and Risk Management}, pages 99--126.
  Springer, 2018.

\bibitem{huang2019adaptive}
L.~Huang and Q.~Zhu.
\newblock Adaptive honeypot engagement through reinforcement learning of
  semi-markov decision processes.
\newblock In {\em International Conference on Decision and Game Theory for
  Security}, pages 196--216. Springer, 2019.

\bibitem{HuangAPT}
L.~Huang and Q.~Zhu.
\newblock A dynamic games approach to proactive defense strategies against
  advanced persistent threats in cyber-physical systems.
\newblock {\em CoRR}, abs/1906.09687, 2019.

\bibitem{huang2020dynamic}
L.~Huang and Q.~Zhu.
\newblock A dynamic games approach to proactive defense strategies against
  advanced persistent threats in cyber-physical systems.
\newblock {\em Computers \& Security}, 89:101660, 2020.

\bibitem{huang2019dynamic}
Y.~Huang, J.~Chen, L.~Huang, and Q.~Zhu.
\newblock Dynamic games for secure and resilient control system design.
\newblock {\em National Science Review}, page to appear, 2020.

\bibitem{huang2019deceptive}
Y.~Huang and Q.~Zhu.
\newblock Deceptive reinforcement learning under adversarial manipulations on
  cost signals.
\newblock {\em arXiv preprint arXiv:1906.10571}, 2019.

\bibitem{jajodia2011moving}
S.~Jajodia, A.~K. Ghosh, V.~Swarup, C.~Wang, and X.~S. Wang.
\newblock {\em Moving target defense: creating asymmetric uncertainty for cyber
  threats}, volume~54.
\newblock Springer Science \& Business Media, 2011.

\bibitem{jajodia2016cyber}
S.~Jajodia, V.~Subrahmanian, V.~Swarup, and C.~Wang.
\newblock {\em Cyber deception}.
\newblock Springer, 2016.

\bibitem{maharjan2013dependable}
S.~Maharjan, Q.~Zhu, Y.~Zhang, S.~Gjessing, and T.~Basar.
\newblock Dependable demand response management in the smart grid: A
  stackelberg game approach.
\newblock {\em IEEE Transactions on Smart Grid}, 4(1):120--132, 2013.

\bibitem{maharjan2015demand}
S.~Maharjan, Q.~Zhu, Y.~Zhang, S.~Gjessing, and T.~Ba{\c{s}}ar.
\newblock Demand response management in the smart grid in a large population
  regime.
\newblock {\em IEEE Transactions on Smart Grid}, 7(1):189--199, 2015.

\bibitem{manshaei2013game}
M.~H. Manshaei, Q.~Zhu, T.~Alpcan, T.~Bac{\c{s}}ar, and J.-P. Hubaux.
\newblock Game theory meets network security and privacy.
\newblock {\em ACM Computing Surveys (CSUR)}, 45(3):25, 2013.

\bibitem{miao2018hybrid}
F.~Miao, Q.~Zhu, M.~Pajic, and G.~J. Pappas.
\newblock A hybrid stochastic game for secure control of cyber-physical
  systems.
\newblock {\em Automatica}, 93:55--63, 2018.

\bibitem{nugraha2020dynamic}
Y.~Nugraha, A.~Cetinkaya, T.~Hayakawa, H.~Ishii, and Q.~Zhu.
\newblock Dynamic resilient network games with applications to multiagent
  consensus.
\newblock {\em IEEE Transactions on Control of Network Systems}, 8(1):246--259,
  2021.

\bibitem{nugraha2019subgame}
Y.~Nugraha, T.~Hayakawa, A.~Cetinkaya, H.~Ishii, and Q.~Zhu.
\newblock Subgame perfect equilibrium analysis for jamming attacks on resilient
  graphs.
\newblock In {\em 2019 American Control Conference (ACC)}, pages 2060--2065.
  IEEE, 2019.

\bibitem{pawlick_TIFS_18}
J.~Pawlick, J.~Chen, and Q.~Zhu.
\newblock {iSTRICT}: An interdependent strategic trust mechanism for the
  cloud-enabled {Internet} of controlled things.
\newblock {\em IEEE Transactions on Information Forensics and Security},
  14(6):1654--1669, 2018.

\bibitem{pawlick2018modeling}
J.~Pawlick, E.~Colbert, and Q.~Zhu.
\newblock Modeling and analysis of leaky deception using signaling games with
  evidence.
\newblock {\em IEEE Transactions on Information Forensics and Security},
  14(7):1871--1886, 2018.

\bibitem{pawlick2019game}
J.~Pawlick, E.~Colbert, and Q.~Zhu.
\newblock A game-theoretic taxonomy and survey of defensive deception for
  cybersecurity and privacy.
\newblock {\em ACM Computing Surveys (CSUR)}, 52(4):82, 2019.

\bibitem{pawlick2015flip}
J.~Pawlick, S.~Farhang, and Q.~Zhu.
\newblock Flip the cloud: Cyber-physical signaling games in the presence of
  advanced persistent threats.
\newblock In {\em International Conference on Decision and Game Theory for
  Security}, pages 289--308. Springer, 2015.

\bibitem{pawlick_mean-field_2017}
J.~Pawlick and Q.~Zhu.
\newblock A {Mean}-{Field} {Stackelberg} {Game} {Approach} for {Obfuscation}
  {Adoption} in {Empirical} {Risk} {Minimization}.
\newblock {\em arXiv preprint arXiv:1706.02693}, 2017.

\bibitem{pawlick2017strategic}
J.~Pawlick and Q.~Zhu.
\newblock Strategic trust in cloud-enabled cyber-physical systems with an
  application to glucose control.
\newblock {\em IEEE Transactions on Information Forensics and Security},
  12(12):2906--2919, 2017.

\bibitem{peng2020distributed}
G.~Peng, J.~Chen, and Q.~Zhu.
\newblock Distributed stabilization of two interdependent markov jump linear
  systems with partial information.
\newblock {\em IEEE Control Systems Letters}, 5(2):713--718, 2020.

\bibitem{RassZhuBook2020}
S.~Rass, S.~Schauer, S.~K\"onig, and Q.~Zhu.
\newblock {\em Cyber-Security in Critical Infrastructures: A Game-Theoretic
  Approach}.
\newblock Advanced Sciences and Technologies for Security Applications.
  Springer, 2020.

\bibitem{rass_gadapt:_2016}
S.~Rass and Q.~Zhu.
\newblock {GADAPT}: a sequential game-theoretic framework for designing
  defense-in-depth strategies against advanced persistent threats.
\newblock In {\em International {Conference} on {Decision} and {Game} {Theory}
  for {Security}}, pages 314--326. Springer International Publishing, 2016.

\bibitem{rieger2019}
C.~Rieger, I.~Ray, Q.~Zhu, and M.~Haney.
\newblock {\em Industrial Control Systems Security and Resiliency: Practice and
  Theory}.
\newblock Advances in Information Security. Springer, 2019.

\bibitem{rinaldi2001identifying}
S.~M. Rinaldi, J.~P. Peerenboom, and T.~K. Kelly.
\newblock Identifying, understanding, and analyzing critical infrastructure
  interdependencies.
\newblock {\em IEEE control systems magazine}, 21(6):11--25, 2001.

\bibitem{song2019performance}
J.~B. Song and Q.~Zhu.
\newblock Performance of dynamic secure routing game.
\newblock In {\em Game Theory for Networking Applications}, pages 37--56.
  Springer, 2019.

\bibitem{spence1973job}
M.~Spence.
\newblock Job market signaling.
\newblock In {\em Uncertainty in economics}, pages 281--306. Elsevier, 1978.

\bibitem{van2013flipit}
M.~Van~Dijk, A.~Juels, A.~Oprea, and R.~L. Rivest.
\newblock Flipit: The game of ``stealthy takeover".
\newblock {\em Journal of Cryptology}, 26(4):655--713, 2013.

\bibitem{xu2015cyber}
Z.~Xu and Q.~Zhu.
\newblock A cyber-physical game framework for secure and resilient multi-agent
  autonomous systems.
\newblock In {\em 2015 54th IEEE Conference on Decision and Control (CDC)},
  pages 5156--5161. IEEE, 2015.

\bibitem{xu2016cross}
Z.~Xu and Q.~Zhu.
\newblock Cross-layer secure cyber-physical control system design for networked
  3d printers.
\newblock In {\em 2016 American Control Conference (ACC)}, pages 1191--1196.
  IEEE, 2016.

\bibitem{xu2017game}
Z.~Xu and Q.~Zhu.
\newblock A game-theoretic approach to secure control of communication-based
  train control systems under jamming attacks.
\newblock In {\em Proceedings of the 1st International Workshop on Safe Control
  of Connected and Autonomous Vehicles}, pages 27--34. ACM, 2017.

\bibitem{xu2018cross}
Z.~Xu and Q.~Zhu.
\newblock Cross-layer secure and resilient control of delay-sensitive networked
  robot operating systems.
\newblock In {\em 2018 IEEE Conference on Control Technology and Applications
  (CCTA)}, pages 1712--1717. IEEE, 2018.

\bibitem{8779673}
H.~Yunhan and Z.~Quanyan.
\newblock A differential game approach to decentralized virus-resistant weight
  adaptation policy over complex networks.
\newblock {\em IEEE Transactions on Control of Network Systems}, pages 1--1,
  2019.

\bibitem{zhang2015secure}
R.~Zhang and Q.~Zhu.
\newblock Secure and resilient distributed machine learning under adversarial
  environments.
\newblock In {\em 2015 18th International Conference on Information Fusion
  (Fusion)}, pages 644--651. IEEE, 2015.

\bibitem{zhang2016attack}
R.~Zhang and Q.~Zhu.
\newblock Attack-aware cyber insurance of interdependent computer networks.
\newblock 2016.

\bibitem{zhang2018game}
R.~Zhang and Q.~Zhu.
\newblock A game-theoretic approach to design secure and resilient distributed
  support vector machines.
\newblock {\em IEEE Transactions on Neural Networks and Learning Systems},
  2018.

\bibitem{zhang2019flipin}
R.~Zhang and Q.~Zhu.
\newblock Flipin: A game-theoretic cyber insurance framework for
  incentive-compatible cyber risk management of internet of things.
\newblock {\em IEEE Transactions on Information Forensics and Security}, 2019.

\bibitem{zhang2017bi}
R.~Zhang, Q.~Zhu, and Y.~Hayel.
\newblock A bi-level game approach to attack-aware cyber insurance of computer
  networks.
\newblock {\em IEEE Journal on Selected Areas in Communications},
  35(3):779--794, 2017.

\bibitem{zhang2017strategic}
T.~Zhang and Q.~Zhu.
\newblock Strategic defense against deceptive civilian gps spoofing of unmanned
  aerial vehicles.
\newblock In {\em International Conference on Decision and Game Theory for
  Security}, pages 213--233. Springer, 2017.

\bibitem{zheng2012dynamic}
J.~Zheng and D.~A. Casta{\~n}{\'o}n.
\newblock Dynamic network interdiction games with imperfect information and
  deception.
\newblock In {\em 2012 IEEE 51st IEEE Conference on Decision and Control
  (CDC)}, pages 7758--7763. IEEE, 2012.

\bibitem{zhu2015game}
Q.~Zhu and T.~Ba\c{s}ar.
\newblock Game-theoretic methods for robustness, security, and resilience of
  cyberphysical control systems: games-in-games principle for optimal
  cross-layer resilient control systems.
\newblock {\em Control Systems, IEEE}, 35(1):46--65, 2015.

\bibitem{zhu2011indices}
Q.~Zhu and T.~Ba{\c{s}}ar.
\newblock Indices of power in optimal ids default configuration: theory and
  examples.
\newblock In {\em Decision and Game Theory for Security}, pages 7--21.
  Springer, 2011.

\bibitem{Quanyan2011ECDC}
Q.~Zhu and T.~Ba{\c{s}}ar.
\newblock Robust and resilient control design for cyber-physical systems with
  an application to power systems.
\newblock In {\em 2011 50th IEEE Conference on Decision and Control and
  European Control Conference}, pages 4066--4071. IEEE, 2011.

\bibitem{zhu2013game}
Q.~Zhu and T.~Ba{\c{s}}ar.
\newblock Game-theoretic approach to feedback-driven multi-stage moving target
  defense.
\newblock In {\em International Conference on Decision and Game Theory for
  Security}, pages 246--263. Springer, 2013.

\bibitem{Quanyan2013CCPS}
Q.~Zhu, L.~Bushnell, and T.~Ba{\c{s}}ar.
\newblock Resilient distributed control of multi-agent cyber-physical systems.
\newblock In {\em Control of Cyber-Physical Systems}, pages 301--316. Springer,
  2013.

\bibitem{zhu2012deceptive}
Q.~Zhu, A.~Clark, R.~Poovendran, and T.~Ba{\c{s}}ar.
\newblock Deceptive routing games.
\newblock In {\em 2012 IEEE 51st IEEE Conference on Decision and Control
  (CDC)}, pages 2704--2711. IEEE, 2012.

\bibitem{zhu2012guidex}
Q.~Zhu, C.~Fung, R.~Boutaba, and T.~Ba{\c{s}}ar.
\newblock Guidex: A game-theoretic incentive-based mechanism for intrusion
  detection networks.
\newblock {\em Selected Areas in Communications, IEEE Journal on},
  30(11):2220--2230, 2012.

\bibitem{zhu2010stochastic}
Q.~Zhu, H.~Li, Z.~Han, and T.~Ba\c{s}ar.
\newblock A stochastic game model for jamming in multi-channel cognitive radio
  systems.
\newblock In {\em 2010 IEEE International Conference on Communications}, pages
  1--6. IEEE, 2010.

\bibitem{zhu2018multi}
Q.~Zhu and S.~Rass.
\newblock On multi-phase and multi-stage game-theoretic modeling of advanced
  persistent threats.
\newblock {\em IEEE Access}, 6:13958--13971, 2018.

\bibitem{zhu2011eavesdropping}
Q.~Zhu, W.~Saad, Z.~Han, H.~V. Poor, and T.~Ba{\c{s}}ar.
\newblock Eavesdropping and jamming in next-generation wireless networks: A
  game-theoretic approach.
\newblock In {\em Military Communications Conference (MILCOM), 2011}, pages
  119--124. IEEE, 2011.

\bibitem{zhu2011dynamic}
Q.~Zhu, J.~B. Song, and T.~Ba{\c{s}}ar.
\newblock Dynamic secure routing game in distributed cognitive radio networks.
\newblock In {\em Global Telecommunications Conference (GLOBECOM 2011), 2011
  IEEE}, pages 1--6. IEEE, 2011.

\bibitem{zhu2010network}
Q.~Zhu, H.~Tembine, and T.~Ba\c{s}ar.
\newblock Network security configurations: A nonzero-sum stochastic game
  approach.
\newblock In {\em American Control Conference (ACC), 2010}, pages 1059--1064.
  IEEE, 2010.

\bibitem{Quanyan2010ACC}
Q.~Zhu, H.~Tembine, and T.~Ba{\c{s}}ar.
\newblock Network security configurations: A nonzero-sum stochastic game
  approach.
\newblock In {\em Proceedings of the 2010 American Control Conference}, pages
  1059--1064. IEEE, 2010.

\bibitem{zhucross}
Q.~Zhu and Z.~Xu.
\newblock {\em Cross-Layer Design for Secure and Resilient Cyber-Physical
  Systems: A Decision and Game Theoretic Approach}.
\newblock Springer Nature, 2020.

\bibitem{zhu2010dynamic}
Q.~Zhu, Z.~Yuan, J.~B. Song, Z.~Han, and T.~Ba\c{s}ar.
\newblock Dynamic interference minimization routing game for on-demand
  cognitive pilot channel.
\newblock In {\em Global Telecommunications Conference (GLOBECOM 2010), 2010
  IEEE}, pages 1--6. IEEE, 2010.

\bibitem{zhu2012interference}
Q.~Zhu, Z.~Yuan, J.~B. Song, Z.~Han, and T.~Ba{\c{s}}ar.
\newblock Interference aware routing game for cognitive radio multi-hop
  networks.
\newblock {\em Selected Areas in Communications, IEEE Journal on},
  30(10):2006--2015, 2012.

\bibitem{Zimmerman-17}
R.~Zimmerman, Q.~Zhu, F.~De~Leon, and Z.~Guo.
\newblock Conceptual modeling framework to integrate resilient and
  interdependent infrastructure in extreme weather.
\newblock {\em Journal of Infrastructure Systems}, 23(4):04017034, 2017.

\bibitem{zimmerman_promoting_2016}
R.~Zimmerman, Q.~Zhu, and C.~Dimitri.
\newblock Promoting resilience for food, energy, and water interdependencies.
\newblock {\em Journal of Environmental Studies and Sciences}, 6(1):50--61,
  2016.

\bibitem{zimmerman2018network}
R.~Zimmerman, Q.~Zhu, and C.~Dimitri.
\newblock A network framework for dynamic models of urban food, energy and
  water systems {(FEWS)}.
\newblock {\em Environmental Progress \& Sustainable Energy}, 37(1):122--131,
  2018.

\end{thebibliography}





\end{document}